\documentclass{ws-procs9x6}
\usepackage{amsmath}
\begin{document}
\title{The DIS($\chi$) Scheme for Heavy Quark Production at Small x}

\author{C.~D. WHITE}

\address{Cavendish Laboratory, J.~J. Thomson Avenue, \\
Cambridge, CB3 0HE, UK, \\ 
E-mail: cdw24@hep.phy.cam.ac.uk}

\maketitle

\abstracts{In order to successfully describe DIS data, one must take heavy quark mass
effects into account. This is often achieved using so called variable
flavour number schemes, in which a parton distribution for the heavy quark
species is defined above a suitable matching scale. At small x, one must
also potentially include high energy corrections to this framework arising from
the BFKL equation. We outline the definition of a variable flavour
scheme which allows such corrections to be consistently implemented
alongside a fixed order QCD expansion. Results of a global fit using this
scheme are presented. We also discuss an extension of the resummation to NLL order.}

\section{Variable Flavour Number Schemes}
In DIS one can produce heavy quarks in the final state via boson gluon fusion. Taking into account all the relevant Feynman diagrams, one may absorb the mass effects into the coefficient functions in a chosen factorisation scheme. This is a {\it fixed flavour number scheme}, and is undesirable at high $Q^2$ in that the coefficients diverge due to collinear singularities. Instead, one may define a heavy quark distribution above a suitable matching scale (here chosen as $Q^2=M^2$), which evolves according to DGLAP equations. This is a so-called {\it variable flavour} description, as $n_f$ changes across $Q^2=M^2$. Equivalence of the FF and VF descriptions imposes a consistency relation between the partons above and below the matching scale\cite{Buza}:
\begin{equation}
\Rightarrow\left(\begin{array}{c}q_{H+}\\g^{(n_f+1)}\\\Sigma^{(n_f+1)}\end{array}\right)=\left(\begin{array}{cc}A_{Hq}&A_{Hg}\\A_{gq}&A_{gg}\\A_{qq}&A_{qg}\end{array}\right)\left(\begin{array}{c}\Sigma^{(n_f)}\\g^{(n_f)}\end{array}\right),
\label{As}
\end{equation}
which defines the heavy matrix elements $\{A_{ij}\}$. This leads to an ambiguity in the VF coefficient functions. For example, at ${O}(\alpha_S)$, one finds:
\begin{equation}
C_{2,H}^{VF(0)}\otimes A_{Hg}^{(1)}+C_{2,g}^{VF(1)}=C_{2,g}^{FF(1)}.
\label{ambiguity}
\end{equation}
Provided this relation is satisfied, one has a valid VF number scheme. However, one is free to shift terms $\sim {O}(M^2/Q^2)$ between the VF coefficients. Each choice corresponds to a different VF number scheme, and this transformation does not change the collinear factorisation scheme. In practice one specifies the placement of $M^2/Q^2$ terms by fixing one of the VF coefficients according to some choice.
\section{Heavy Quark Production at Small $x$}
Our aim is find a VF number scheme that has the same definition at fixed order as it does in the high energy expansion (small $x$). It is convenient to adopt the DIS factorisation scheme for this purpose, as then the LL impact factors coupling the virtual photon to the BFKL gluon ladder via a massless quark pair are easily interpreted in terms of the splitting function $P_{qg}$ and longitudinal coefficient $C_{Lg}$. There is no mixing between $C_{2g}$ and $P_{qg}$ as the former quantity is zero in this scheme. One must also choose the placement of $M^2/Q^2$ terms. Previously, the ACOT group has proposed the identification $C_{2,H}^{\overline{\text{MS}}(0)VF}=\delta(1-x')$ (the ACOT($\chi$) scheme\cite{ACOTchi}), where $x'$ is the scaled Bjorken variable $x'=x(1+4M^2/Q^2)$ taking into account the kinematic constraint at the heavy quark vertex. We propose the same identification in the DIS scheme n.b. $C_{2,H}^{VF}=\delta(1-x')$ to all orders. This then specifies completely $C_{2,g}^{VF}$ and $C_{2,q}^{VF}$ and we call this the DIS($\chi$) scheme by analogy with ACOT. It is then easy to interpret the heavy quark impact factors at small $x$, and in double Mellin space one finds the resummed quantities:
\begin{align}
A_{Hg}(\gamma,N,Q^2/M^2)&=h_2(\gamma,N,Q^2/M^2)|_{\frac{Q^2}{M^2}\rightarrow\infty};\label{AHg}\\
C_{2,g}^{VF}&=h_2(\gamma,N,Q^2/M^2)-A_{Hg}(\gamma,N,Q^2/M^2);\label{c2gvf}\\
C_{L,g}^{VF}&=C_{L,g}^{FF}=h_L(\gamma,N,Q^2/M^2),\label{clgvf}
\end{align}
with $N$ Mellin conjugate to $x$ and $\gamma$ conjugate to $Q^2/\Lambda^2$. Results in $x$ and $Q^2$ space depend upon how one solves the BFKL equation. 
\section{A LL Resummed Global Fit with Running Coupling}
\begin{figure}
\begin{center}
\scalebox{0.32}{\includegraphics{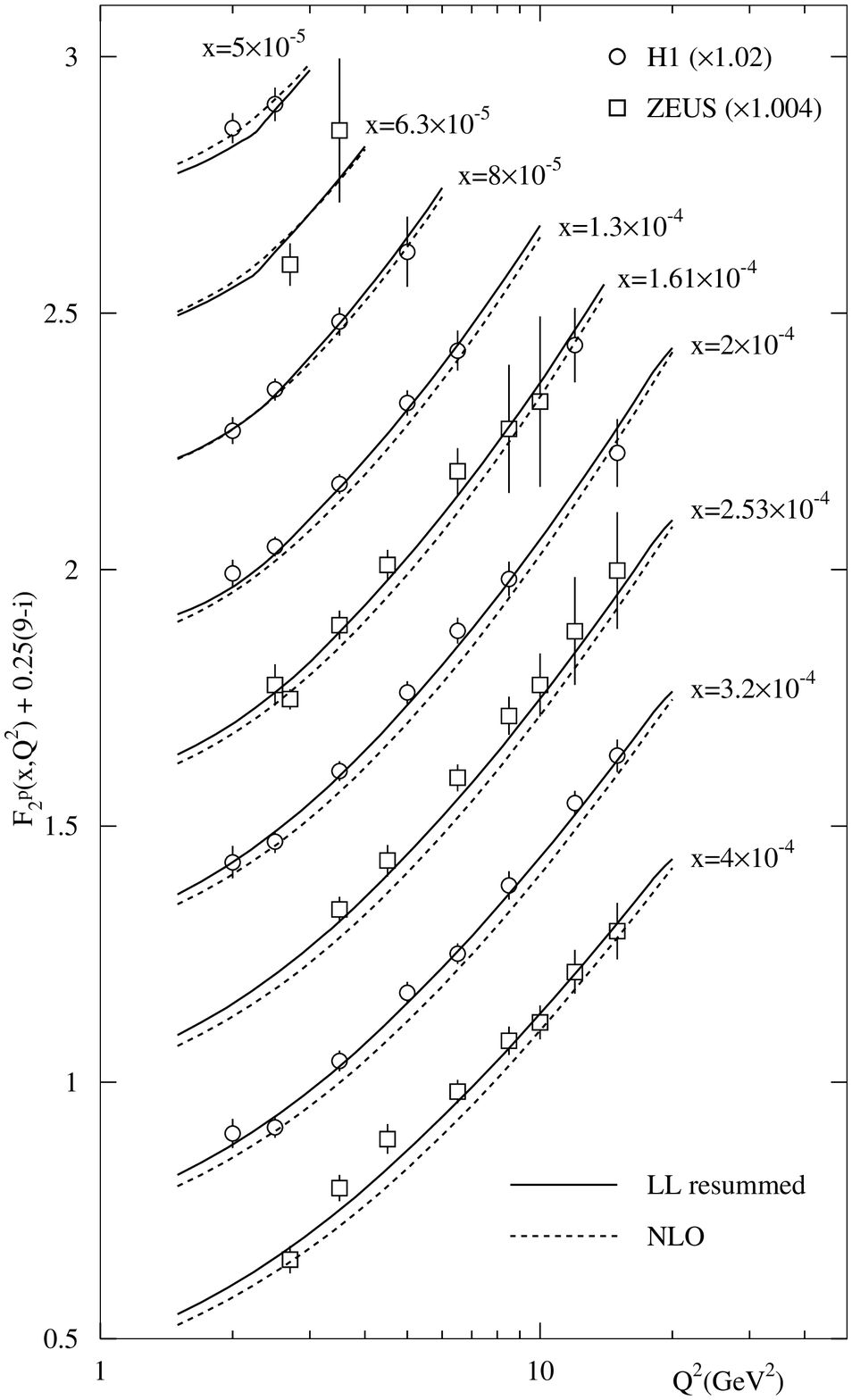}}
\scalebox{0.32}{\includegraphics{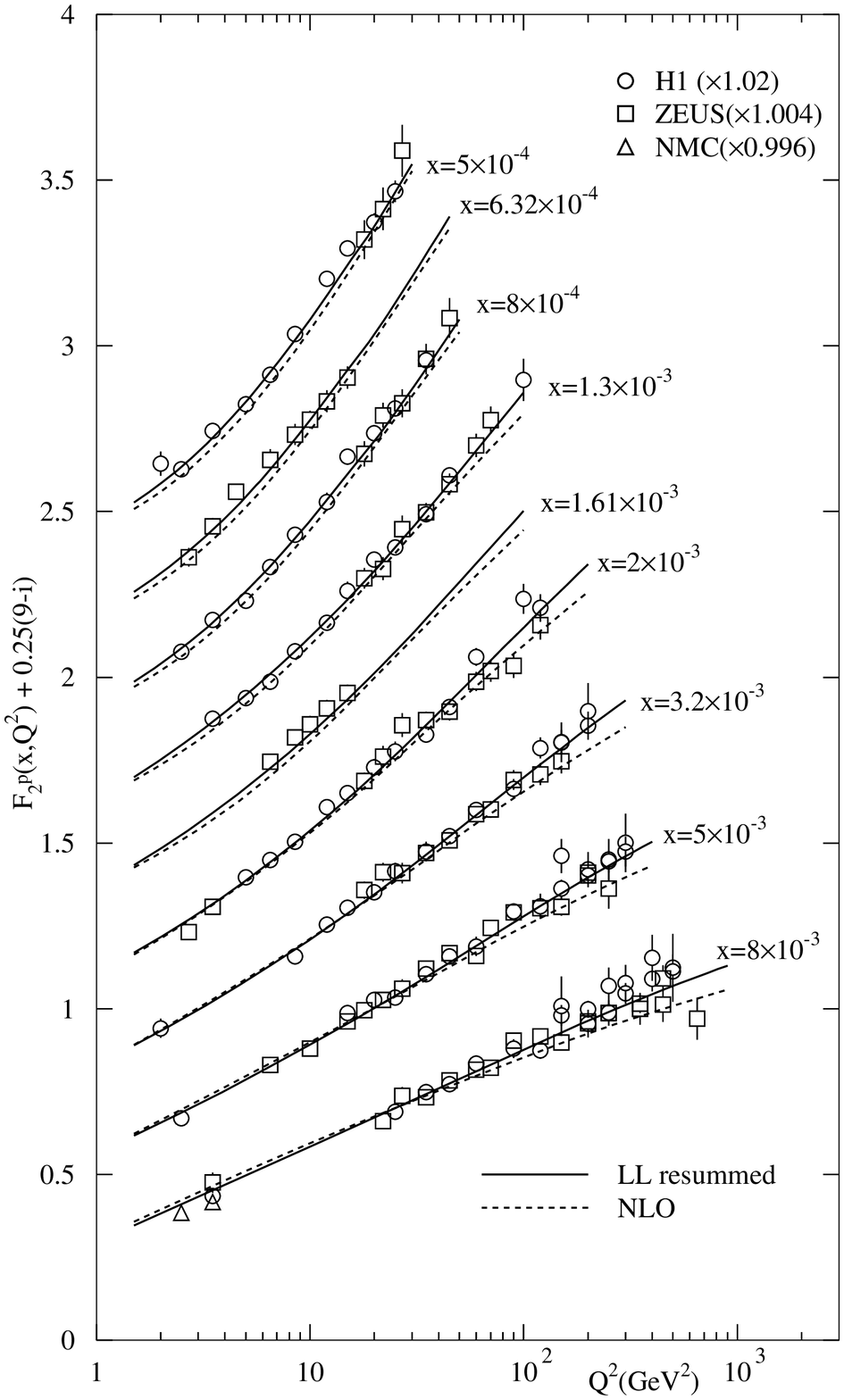}}
\caption{Comparison of a LL resummed fit (with running coupling) to data from HERA and NMC. Also shown is a NLO fit result.}
\label{data}
\end{center}
\end{figure}

We implemented the DIS($\chi$) scheme in a LO global fit with LL resummations\cite{TW}, where the solution of the BFKL equation also included the running coupling \cite{Thorne01}. Although this latter effect is strictly subleading, the resulting resummed coefficients show a considerably softer divergence at small $x$. The resummed prediction is compared with data alongside a standard NLO prediction from a global fit in figure \ref{data}.
\vspace{-1cm}
The resummed prediction is excellent at small $x$, whereas the NLO description starts to fail because its slope is insufficient as $Q^2$ increases. However, a problem with the resummed fit (not shown) is that the predicted gluon and sea distributions (hence $F_L$ and jet cross-sections) are too low in the moderate $x$ region. This indicates a need for NLO as well as NLL corrections, to suppress resummation effects to lower $x$ values.
\section{Toward a NLL Analysis with Exact Gluon Kinematics}
\begin{figure}
\begin{center}
\scalebox{0.6}{\includegraphics{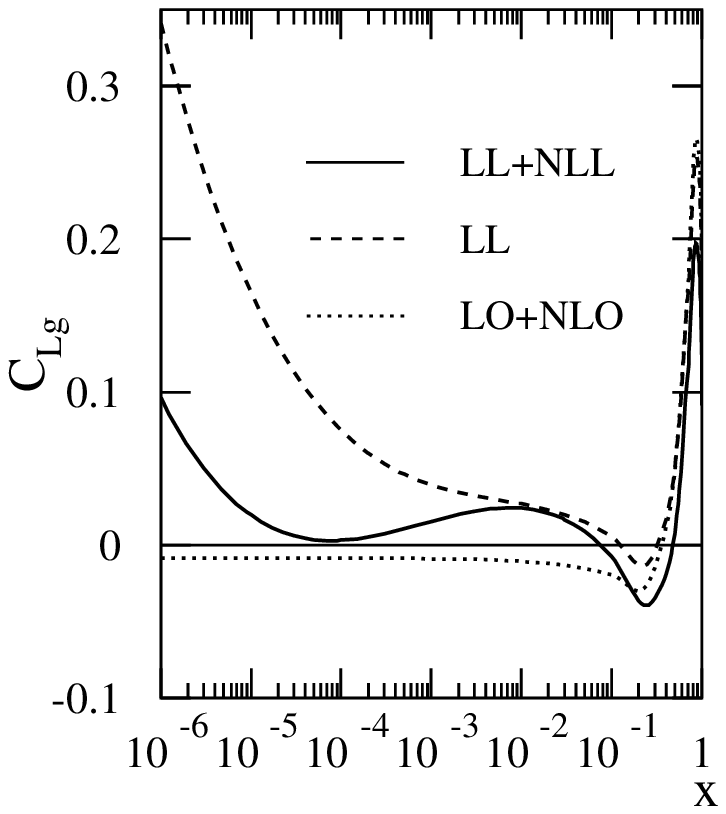}}
\scalebox{0.6}{\includegraphics{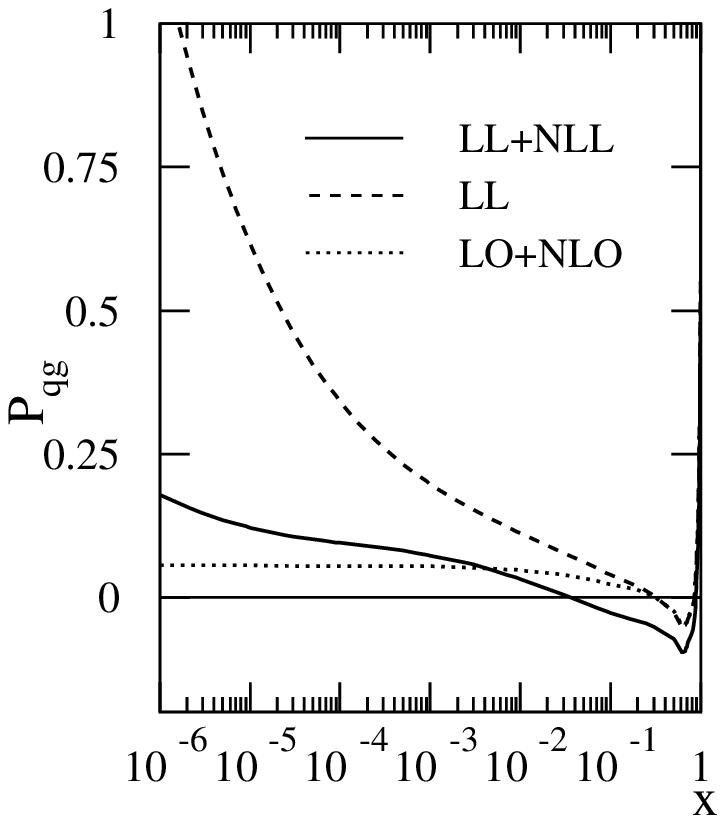}}
\caption{Preliminary result for $P_{qg}$ with NLL and running coupling corrections. Also shown are LL (with running coupling) and fixed order results.}
\label{P_qg}
\end{center}
\end{figure}

In calculating the impact factors coupling the virtual photon to the BFKL gluon ladder, one assumes that the longitudinal momentum fraction of the gluon at the top of the ladder is the same as Bjorken $x$ (corresponding to the bottom of the ladder). This is not true beyond LL order, and the impact factors with the exact gluon kinematics have been previously calculated \cite{Peschanski}. We have shown that they seem to contain most of the NLL correction to the impact factor \cite{TW2}, and thus pave the way for an approximate NLL analysis of scattering data. A preliminary result for the NLL corrected $P_{qg}$ is shown in figure \ref{P_qg}. The NLL corrections soften the low $x$ divergence below the LL prediction and delay the onset of asymptotic behaviour to lower $x$ values. This is hopefully what is needed to obtain a good description of scattering data over the whole of the accessible $x$ range!\\
\vspace{-2cm}

{\it Acknowledgements}. This work is the result of collaboration with my graduate supervisor Robert Thorne. I am grateful to PPARC for a research studentship.

\end{document}